\documentclass[useAMS,usenatbib]{mn2e}
\usepackage{epsfig,graphics,graphicx,bm,amssymb}

\begin{document}

\title{Mapping the $\nu_\odot$ Secular Resonance for Retrograde Irregular Satellites}

\author[J. Correa Otto, A.M. Leiva, C.A. Giuppone and C. Beaug\'e]
{J. Correa Otto$^*$, A.M. Leiva, C.A. Giuppone and C. Beaug\'e \\
Observatorio Astron\'omico, Universidad Nacional de C\'ordoba, Laprida 854, (X5000BGR) C\'ordoba, Argentina \\
$^*$ e-mail: jorgecorrea@oac.uncor.edu}

\maketitle

\begin{abstract}
Constructing dynamical maps from the filtered output of numerical integrations,
we analyze the structure of the $\nu_\odot$ secular resonance for fictitious
irregular satellites in retrograde orbits. This commensurability is associated
to the secular angle $\theta = \varpi - \varpi_\odot$, where $\varpi$ is the
longitude of pericenter of the satellite and $\varpi_\odot$ corresponds to the
(fixed) planetocentric orbit of the Sun. Our study is performed in the
restricted three-body problem, where the satellites are considered as massless
particles around a massive planet and perturbed by the Sun.

Depending on the initial conditions, the resonance presents a diversity of
possible resonant modes, including librations of $\theta$ around zero (as found
for Sinope and Pasiphae) or $180$ degrees, as well as asymmetric librations
(e.g. Narvi). Symmetric modes are present in all giant planets, although each
regime appears restricted to certain values of the satellite inclination.
Asymmetric solutions, on the other hand, seem absent around Neptune due to its almost circular heliocentric orbit.

Simulating the effects of a smooth orbital migration on the satellite, we find
that the resonance lock is preserved as long as the induced change in semimajor
axis is much slower compared to the period of the resonant angle (adiabatic
limit). However, the librational mode may vary during the process, switching
between symmetric and asymmetric oscillations.

Finally, we present a simple scaling transformation that allows to estimate the
resonant structure around any giant planet from the results calculated around a
single primary mass.
\end{abstract}


\section{Introduction}

Both Jupiter and Saturn are rich in irregular satellites. Jupiter has at least
54 moons in this group, while Saturn contains more than 30 members. A smaller
population is also observed around Uranus and Neptune, although it is not clear
whether this is intrinsic or due to observational bias.

Irregular satellites suffer strong perturbations from the Sun which
significantly affect their orbits evolution. Whipple \& Shelus (1993)
numerically integrated the orbit of the Jovian moon Pasiphae for $10^5$ years
under the effects of all planets, and found that its longitude of pericenter
$\varpi$ is locked in a secular resonance where $\varpi - \varpi_{\rm Jup}$
librates around $180^\circ$. Here $\varpi_{\rm Jup}$ is the longitude of
pericenter of the heliocentric orbit of Jupiter. Since $\varpi_{\rm Jup} =
\varpi_\odot + \pi$, where $\varpi_\odot$ is the Jupiter-centric longitude of
the pericenter of the Sun, this implies that $\varpi - \varpi_{\odot}$
oscillates around zero. By analogy from asteroidal motion, we will call this
the $\nu_5$ secular resonance for satellite orbits.

In the same year, Saha \& Tremaine (1993) found that the Jovian moon Sinope
also displays a similar resonance lock in the $\nu_5$ resonance, altough in
this case the motion alternates between libration and circulation on timescales
of the order of $10^5-10^6$ years. Extending the orbital evolution for $5
\times 10^7$ years, Nesvorn\'y et al. (2003) showed that Pasiphae's libration
is also temporary, and will switch to a circulation of the resonant angle in
approximately $3 \times 10^7$ years.

\'Cuk \& Burns (2004) performed a detailed search for other irregular moons in
secular resonances, finding that the prograde Saturn moon Siarnaq appears to
the trapped in a pericenter secular resonance with Saturn (i.e. $\nu_6$), also
displaying intermittent libration, where the resonant angle oscillates around
$180^\circ$. While the Uranus irregular satellite Stephano is not in
libration, the angle $\varpi - \varpi_{\odot}$ shows a quasi-resonant behavior
with a very long-period circulation. 

Finally, Beaug\'e \& Nesvorn\'y (2007) found that although the retrograde Saturn moon Narvi is not currently in resonant motion, long-term simulations show a future libration in the $\nu_6$ secular resonance. Contrary to previous
cases, here the resonant angle displays libration around values different from zero or $180^\circ$, in what appears to be an asymmetric libration point. 
Since all these resonant configurations, independently of the planetary mass, involve the relative behavior of the longitude of pericenter of the satellite and that of the planetocentric orbit of the Sun, in all future references we will denote this commensurability as $\nu_\odot$. 

At present it is not clear whether the observed resonant population in the outer planets is evidence of a past smooth orbital migration of the satellites, or whether it is simply due to chance. Beaug\'e \& Nesvorn\'y (2007) found that stability criteria alone yield satellite populations close to $\nu_\odot$, although the proximity of some individual moons (particularly Pasiphae and Sinope) appear sufficiently detached from a random distribution to be statistically
significant.

One of the problems in understanding a possible relationship between the
$\nu_\odot$ resonance and the past evolution of the irregular satellites, is the lack of a detailed analysis of the resonant structure. As will be shown in Section 2, analytical models, even of high-order, are not sufficiently precise and are unable to reproduce the main characteristics of the secular commensurability. Semi-analytical models, where the secular perturbations are evaluated by numerical averaging of the exact Hamiltonian, also suffer the same limitations since they are equivalent to a first-order theory. For this reason, so far most dynamical studies have been restricted to numerical integrations of individual orbits.

In this paper we present a numerical study of the structure of the $\nu_\odot$
resonance for retrograde orbits. The resonant behavior is obtained applying two
filters on the numerical output: the first is constructed to eliminate the
short-period variations (frequencies comparable with the mean motions). The
resulting data is then filtered once again to reduce the effects caused by the
non-resonant secular angular variable. The results are displayed as dynamical
maps showing the location, type and extension of the librational domains for
several values of the integrals of motion.

The details of the numerical method are outlined in Section 3, which includes
an application to the region of the phase space in the vicinity of the Jovian
moon Sinope. Section 4 shows the structure of the $\nu_\odot$ resonance in the
Saturn system. In both cases the maps are constructed for a single value of the
proper semimajor axis. Section 5 discussed the effects of an ad-hoc migration
acting of the satellites orbits, and its effects on the resonant motion. Next,
in Section 6 we present a simple scaling law that allows us to relate the
resonant structure for any planetary mass. Finally, the application the real
satellites are briefly discussed in Section 7, while concluding remarks close
the paper in Section 8.

\section{Analytical Models for the $\nu_\odot$ Resonance}

Suppose a fictitious satellite in orbit around a giant planet $m_p$ perturbed by the Sun. Let $a$ be the satellite's planetocentric semimayor axis, $e$ its eccentricity, $i$ its inclination with respect to the Laplace plane, $M$ the mean anomaly, $\omega$ the argument of pericenter and $\Omega$ the longitude of the node. Planetocentric orbital elements of the Sun will be identified by the index $\odot$. Since we will adopt a Hamiltonian approach, it is usefull to introduce the modified Delaunay canonical variables:
\begin{equation}
\label{eq0}
\begin{array}{lll}
L = \sqrt{\mu a}  &;& M \\ 
G = L\sqrt{1-e^2} &;& \omega \\
H = G \cos{i} &;& \Omega .
\end{array}
\end{equation}
The complete Hamiltonian function can then be written as $F=F_0(L,\Lambda) + R$, where $F_0$ is the two-body contribution ($\Lambda$ is the canonical momenta associated to $M_\odot$) and $R=R(L,G,H,M,\omega,\Omega,M_1)$ denotes the disturbing function stemming from the Solar gravitational effects. $R$ also depends on the orbital elements of the Sun ($e_{\odot}$, $i_{\odot}$, $\varpi_{\odot}$ and $\Omega_{\odot}$), which are considered constant, as well as on the planetary (i.e. central) mass $m_p$. 

Yokoyama et al. (2003) presented an analytical model for the secular behavior of irregular satellites in the restricted three-body problem, and applied it to the Jovian satellite system. It is based on Kaula's (1962) expansion of the disturbing function, truncated at fourth order in $\alpha = a/a_{\odot}$ and eccentricities and inclinations. The averaging over the short-period terms, associated to the mean anomalies of both bodies, is done to first order. We will denote as $F_1$ the resulting expression of the secular Hamiltonian. 

Although we will not write the expression explicitly, functionally it can be written as $F_1=F_1(G^*,H^*,\omega^*,\Omega^*;\alpha^*)$, where the new (starred) variables are the mean elements, which must not be mistaken with the original osculating variables. the quantity $\alpha^*$ is related to the ``proper'' Delaunay momenta $L^*$ by the equation $\alpha^* = {L^*}^2/\mu a_{\odot}$. 

To study a given secular resonance, we must perform a canonical transformation to resonant variables. The $\nu_\odot$ commensurability is defined by the angle $\theta = \varpi - \varpi_{\odot}$; however, the relationship between the argument of pericenter $\omega$ and the longitude of pericenter $\varpi$ is different when considering direct or retrograde orbits. Since most of the real irregular satellites in the vicinity of this resonance are retrograde, we will assume $i>90^\circ$, for which $\varpi = \Omega - \omega$. 

Our set of resonant variables will be $(I_1,I_2,\phi,Q)$, where $\phi=\Omega^* - \omega^*$ will be a slow angle related to $\theta$, while $Q$ is a fast angle. This can be chosen among any linear combination of the secular angles with the only condition being that it be independent of $\phi$. For the present work we will choose $Q=\Omega^*$, and check that the precessional frequency of the longitude of the node is higher than the frequency of $\phi$. Having specified the angles, the expressions for the corresponding canonical momenta can be easily found. The complete transformation from the mean Delaunay variables to the resonant counterparts are given by:
\begin{equation}
\label{eq1}
\begin{array}{lll}
I_1 = -G^*      &;& \phi = \Omega^* - \omega^* \\ 
I_2 = G^* + H^* &;& Q = \Omega^* .
\end{array}
\end{equation}
The resonant angle would then be $\theta = \phi - \varpi_{\odot}$. 

The transformation of the secular Hamiltonian to resonant variables is straightforward, yielding $F_1(I_1,I_2,\theta,Q)$. Since $Q$ is a fast variable, we can perform a first-order averaging of the Hamiltonian over this angle. This averaging implies a new canonical transformation to new resonant variables $(I_1^*,I_2^*,\theta^*,Q^*)$ such that the transformed Hamiltonian $F_1^*=F_1^*(I_1^*,\theta^*;I_2^*)$ and $Q^*$ is cyclic. The associated momenta $I_2^*$ is then a new constant of motion, and the system is reduced to a single-degree of freedom model for the $\nu_\odot$ resonance in the canonical pair $(I_1^*,\theta^*)$.

Equation (10) of Yokoyama et al. (2003) gives explicit expressions for $F_1^*$, that can be succinctly written as:
\begin{equation}
\label{eq2}
F_1^* = A_0 + A_1 \cos{(\theta^*)} + A_2 \cos{(2\theta^*)},
\end{equation}
where the coefficients $A_k=A_k(I_1^*;L^*,I_2^*)$ are function of the resonant momenta $I_1^*$ and parametrized by the constants $L^*,I_2^*$ and the orbital elements of the Sun. Different initial conditions will give different values of the proper elements $L^*,I_2^*$. 

\subsection{The Secular Dynamics of Sinope}

\begin{figure}
\begin{center}
\epsfig{figure=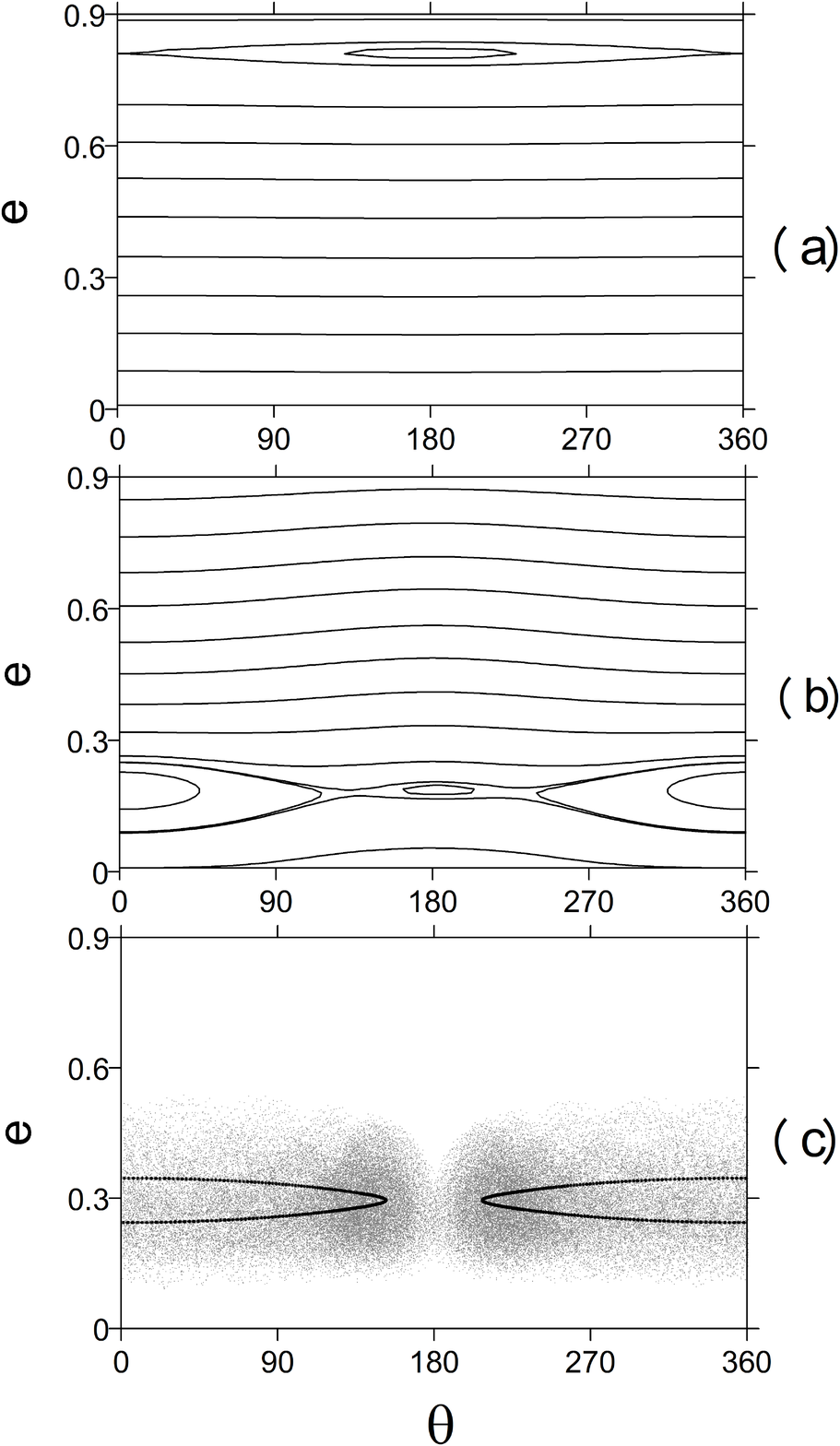,height=14cm,angle=0,clip=}
\caption{{\bf (a).} Level curves of constant Hamiltonian $F_1^*$ for the Jovian  $\nu_\odot$ resonance, in the plane $(e^*,\theta^*)$, using a first-order averaging over short-period terms. The adopted values for the integrals of motion are $L^*=2.115 \times 10^{-4}$,$I_2^*=1.575 \times 10^{-5}$, in units of Solar mass, AU and day. These values correspond to the Jovian satellite Sinope. {\bf (b).} Same as before, but for the third-order Hamiltonian $F_3^*$. {\bf (c).} Numerical simulation of {\it Sinope} (three-body problem) over $10^5$ years. Gray dots show the evolution of the osculating elements, while black dots correspond to an output filtered to eliminate all periodic variations with period smaller than $100$ years.}
\label{fig1}
\end{center}
\end{figure}

As an example, Figure \ref{fig1}(a) shows the level curves of constant $F_1^*$ in the plane $(\theta^*,e^*)$, adopting Jupiter as the central body (i.e. $m_p$). The quantity $e^*=e^*(I_1^*;L^*,I_2^*)$ is the so-called mean-mean eccentricity, which can be obtained from the canonical momenta $I_1^*$ assuming that the values of $L^*,I_2^*$ are constant for all initial conditions. The adopted values for these parameters are close to the irregular satellite Sinope, and were calculated from a numerical simulation of the exact equations (three-body problem) over a timespan of $10^5$ years. 

For these constants, the phase plane of $F_1^*$ shows a resonance region centered approximately at $e^* \simeq 0.85$ corresponding to a libration of $\theta^*$ around $180^\circ$. The width of the libration zone is approximately $\Delta e \sim 0.05$. All initial conditions with lower values of the eccentricity correspond to circulations. This structure contrasts significantly with the Figure \ref{fig1}(c) which shows the real dynamical behavior of Sinope obtained with the same numerical simulation described previously. The values of $(\theta^*,e^*)$ were calculated using a low pass FIR digital filter (Carpino et al. 1987) to eliminate all periodic variations with period smaller than $100$ years. This includes variations in the orbital elements due to both the mean anomalies and $\Omega$. For comparison, in gray dots we also show the evolution of the osculating elements. A comparison between both figures shows that the analytical model $F_1^*$ fails to reproduce the correct dynamics of the system, predicting that Sinope should be far from the resonance domain displaying a circulation of $\theta^*$. 

This imprecision is not restricted to the structure of the phase space in vicinity of this satellite, but is a general flaw of the averaging process used to obtain the analytical model. As shown by \'Cuk \& Burns (2004), the averaging over short period terms must be performed to order higher than unity, to include the effect of second and third order secular terms such as evection. These effects cause significant variations in the secular frequencies, thus affecting both the long-terms evolution of individual bodies, calculations of including proper elements, as well as the location of secular resonances in the domain of irregular satellites. 

With this in mind, Beaug\'e et al. (2006) developed a high order analytical model of the secular Hamiltonian for irregular satellites, where the averaging process over the mean anomalies was extended to third order. The new Hamiltonian function, which we will refer to as $F_3=F_3(G^*,H^*,\omega^*,\Omega^*;\alpha^*)$ was used to calculate the proper elements of all irregular satellites of the outer planets and estimate their proximity to different secular resonances (Beaug\'e \&  Nesvorn\'y 2007). 

From the analytical expression of $F_3$, we can now construct a higher order model for the $\nu_\odot$ resonance in the same manner as before, and plot the level curves of $F_3^*$ in the vicinity of Sinope. Results are shown in Figure \ref{fig1}(b) and show a significant improvement with respect to $F_1$. The resonance domain is now much closer to the actual orbit of the Jovian satellite, and the main libration island is now (correctly) associated to an oscillation around $\theta^*=0$. However, some important discrepancies still remain. The center of the libration region is still not correct, and the analytical model now predicts a smaller eccentricity than the numerical simulation. More important, $F_3^*$ shows an secondary libration island around $\theta^*=180^\circ$ which in fact does not exist. As shown in Figure 16 of Beaug\'e \& Nesvorn\'y (2007) the size of this second libration mode increases for larger values of $I_2^*$, becoming an important characteristic of the phase space for librations at high eccentricities.

\begin{figure}
\begin{center}
\epsfig{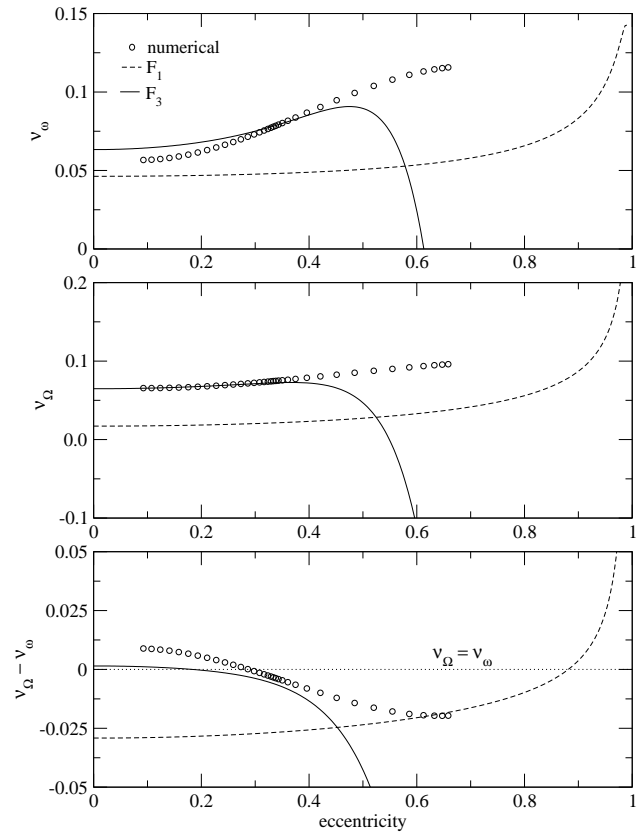}
\caption{Frequencies of the secular angles $\omega^*$, $\Omega^*$ and $\phi = \Omega^* - \omega^*$, obtained with both analytical Hamiltonians (continuous and dashed curves). Open circles show numerical results calculated from a Fourier analysis of filtered data. All initial conditions correspond to the same values of $L^*$ and $I_2^*$ as in figure \ref{fig1}.}
\label{fig2}
\end{center}
\end{figure}

\subsection{Estimation of the Secular Frequencies}

The top and middles plots of Figure \ref{fig2} shows the frequencies of the secular angles $\omega^*$, $\Omega^*$ (denoted by $\nu_\omega$ and $\nu_\Omega$, respectively) as a function of the proper eccentricity $e^*$, with initial conditions corresponding to the same values of the integrals $L^*$ and $I_2^*$ as before. Dashed curves show the results obtained using the analytical Hamiltonian $F_1^*$ restricted to those terms which depend only on the canonical momenta. Frequencies calculated with the third-order function $F_3^*$ are shown in continuous lines. For comparison, we also estimated the same frequencies numerically (open circles), Fourier analyzing the filtered output of a series of simulations for $10^5$ years. 

For both $\omega^*$ and $\Omega^*$ there is a very marked offset between the real frequencies and the results obtained through $F_1^*$, and the analytical estimates systematically yield lower values. A better agreement is noted using $F_3^*$, especially for the precessional frequency of the node, although a larger discrepancy is noted for the argument of pericenter. It is also important to recall that both analytical models were developed using Kaula's expansion of the disturbing function, whose convergence is only guaranteed for eccentricities below $\sim 0.6$ (e.g. Wintner 1941). This seems to be the cause of the increasing imprecision of $F_3^*$ for values of $e^*$ approaching this limit, but also implies that the results obtained with the simpler model $F_1^*$ are also undependable for $e^*> 0.6$. 

The bottom graph of Figure \ref{fig2} shows the values of $\nu_\Omega - \nu_\omega$ for the same initial conditions. Zero values (dotted line) correspond to the exact $\nu_\odot$ resonance, since we are considering constant longitude of pericenter for Jupiter's orbit. this plot shows how the imprecision in the secular frequencies affect the resonance location. For $F_1^*$ the exact resonance occurs for almost parabolic orbits, well beyond the convergence limit of the expansion of the disturbing function, and thus its very existence appears unreliable. Once again better agreement is noted using $F_3^*$, although a significant error is still in evidence. Notwithstanding, since the continuous curve is very shallow for low eccentricities, even small differences in the model can cause important displacements in the location of the exact resonance.

\section{The Numerical Model}

Outside secular resonances, analytical models constitute adequate tools for the long-term dynamics of irregular satellites, and can be used to obtain fairly precise proper elements for real bodies (Beaug\'e \& Nesvorn\'y 2007). They are also able to reproduce the structure and location of the Lidov-Kozai resonance (Lidov 1961, Kozai 1962) with good precision. However, even high order analytical models seem insufficient for the task of mapping the structure of other secular resonances, such as the $\nu_\odot$ commensurability. Not only may the location of the resonance domain may be imprecise, but the libration centers and secondary modes predicted by the models may be unreliable.

In this section we will present a series of dynamical maps of the Jovian $\nu_\odot$ resonance, obtained numerically by filtering the outputs of a series of long-term simulations of fictitious particles. Each map consists of the dynamical evolution of a set of initial conditions with predefined values of the integrals $({\bar L}^*,{\bar I}_2^*)$. The main advantage of this procedure over analytical expansions is that we are no longer restricted to low to moderate eccentricities, and there are no approximations in the modeling of the gravitational interactions. As before we will work in the realm of the restricted three-body problem and assume a fixed orbit for the planetocentric motion of the Sun. 

For each map we begin with a set of initial conditions in osculating orbital elements for the particle plus the mean anomaly of the Sun $(a,e,i,M,M_{\odot},\omega,\Omega)$. Each orbit is integrated for a time span of $10^5$ years using a Burlisch-Stoer based N-body code. The output is subsequently converted to canonical resonant variables $(L,I_1,I_2,M,M_{\odot},\phi,Q)$ and filtered to eliminate the variations associated to the mean anomalies $M,M_{\odot}$, as well those corresponding to the fast variable $Q$. Although the orbital period of both bodies are well known beforehand, the frequency of $Q$ may depend with the initial condition and must be monitored closely. Thus, the decimation and length of the filter may change in each simulation. However, we have found that a cutoff period of $200$ years and a filter of $1600$ data points gave very precise results in most cases. 

The output of the filtering is a numerical approximation to the transformed resonant variables $(I_1^*,\phi^*;L^*,I_2^*)$, analogous to those obtained with an analytical model, with one important difference. Although the digital filtering can be considered analogous to an averaging of the short-period angles, it does not reduce the number of degrees of freedom of the system. Consequently, any chaoticity in the original solution will be preserved in the filtered data. This implies that a regular (non-chaotic) solution will give a one dimensional curve in the $(I_1^*,\phi^*)$ plane, while a stochastic trajectory will cover a two-dimensional region. 

To construct the map, each set of initial conditions in osculating variables must yield the same values of $(L^*,I_2^*)=({\bar L}^*,{\bar I}_2^*)$. From perturbation theory we know that
\begin{eqnarray}
\label{eq3}
L^* &=& L + \frac{\partial}{\partial M} \chi = f_1(a,e,i,M,M_{\odot},\omega,\Omega)  \\
I_2^* &=& I_2 + \frac{\partial}{\partial Q} \chi = 
f_2(a,e,i,M,M_{\odot},\omega,\Omega) \nonumber,
\end{eqnarray}
where $\chi(L,I_1,I_2,M,M_{\odot},\phi,Q)$ is the generating function written in osculating variables evaluated at the initial conditions. Analytical models allow explicit construction of function $\chi$, which can be inverted to give the necessary values of the osculating elements. Since numerical simulations do not give $f_i$, the inversion of expressions (\ref{eq3}) must be done by iterations.

The idea is to introduce small deviations in the osculating variables 
$a \rightarrow a + \Delta a$ and $i \rightarrow i + \Delta i$ and leave all the other initial conditions intact. Introducing these expressions in (\ref{eq3}) and expanding $f_i$ in a first-order Taylor around the original values, we obtain:
\begin{eqnarray}
\label{eq4}
L^* + \Delta L^* &=& L^* + \frac{\partial f_1}{\partial a} \Delta a + 
			   \frac{\partial f_1}{\partial i} \Delta i    \\
I_2^* + \Delta I_2^* &=& I_2^* + \frac{\partial f_2}{\partial a} \Delta a + 
			   \frac{\partial f_2}{\partial i} \Delta i \nonumber,
\end{eqnarray}
where the partial derivatives of $f_i$ are evaluated at the initial orbital set. Equating $L^* + \Delta L^* = {\bar L}^*$ and $I_2^* + \Delta I_2^* = {\bar I}_2^*$, equations (\ref{eq4}) can be inverted to give the necessary values of $\Delta a$ and $\Delta i$. The partial derivatives can be estimated numerically using finite differences, calculating how the integrals vary with very small deviations in each of the orbital elements around the original values. This procedure is not very precise, but can be used a successive approximations to give the desired solution. In general, a few iterations are sufficient to give values with relative precision of the order of 
$\Delta L^*/{\bar L}^* , \Delta I_2^*/{\bar I}_2^* \sim 10^{-4}$. 

\begin{figure}
\begin{center}
\epsfig{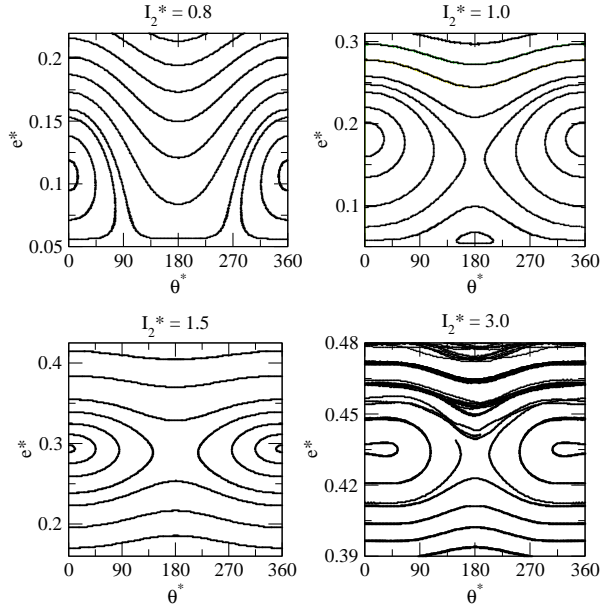}
\caption{Filtered output, in the $(e^*,\theta^*)$ plane, of a series of numerical integrations with initial conditions leading to $a^* = 0.16$ AU ($L^*=2.115 \times 10^{-4}$) and four different values of $I_2^*$ (in units of $10^{-5}$). Each plot is roughly centered on the libration domain of the Jovian $\nu_\odot$ resonance. Chaotic motion is noticeable in the lower right-hand plot, especially for values of $e^*$ larger than the center of libration.}
\label{fig3}
\end{center}
\end{figure}

\subsection{Resonance Maps Around Sinope}

\begin{figure}
\begin{center}
\epsfig{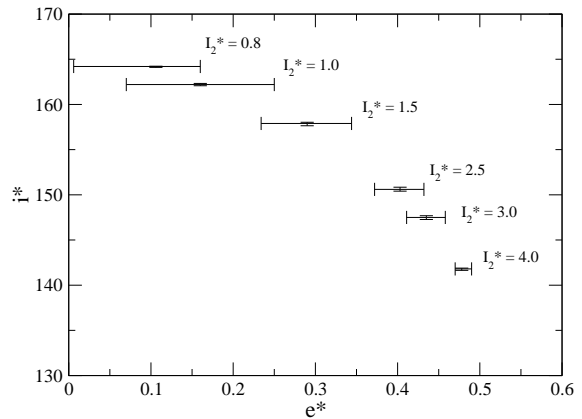}
\caption{Values of $i^*$ and $e^*$ corresponding to the center of the libration island for different values of $I_2^*$. Error bars indicate the libration width in each orbital element.}
\label{fig4}
\end{center}
\end{figure}

Figure \ref{fig3} shows several dynamical maps constructed for $L^*=2.115 \times 10^{-4}$ (corresponding to $a^*=0.16$ AU) and four different values of $I_2^*$. For simplification purposes, all numerical values of $I_2^*$ will be given in units of $10^{-5}$. The lower left-hand frame roughly corresponds the initial conditions of Sinope. Instead of plotting $I_1^*$ as function of the resonant angle $\theta^*$, we converted the canonical momenta to equivalent mean-mean eccentricities $e^*$ and inclinations $i^*$ according to the relationships:
\begin{eqnarray}
\label{eq5}
I_1^* &=& -G^* = -L^* \sqrt{1 - {e^*}^2}  \\
I_2^* &=& G^* + H^* = I_1^*( 1 + \cos{i^*}) \nonumber .
\end{eqnarray}
In all the plots we notice a main libration island centered around $\theta^*= 0$, altough the map for $I_2^*=1.0$ also shows a small libration region centered at $\theta^*=180^\circ$. The values of $i^*$ and $e^*$ of the center of each main libration island are shown Figure \ref{fig4}, where the error bars indicate the libration width in each orbital element. The libration zone appears extensive in the eccentricity, although it decreases in size for larger values of $I_2^*$. However, the corresponding size in $i^*$ is very small, indicating that the $\nu_\odot$ commensurability appears very restricted in the inclination.

\begin{figure}
\begin{center}
\epsfig{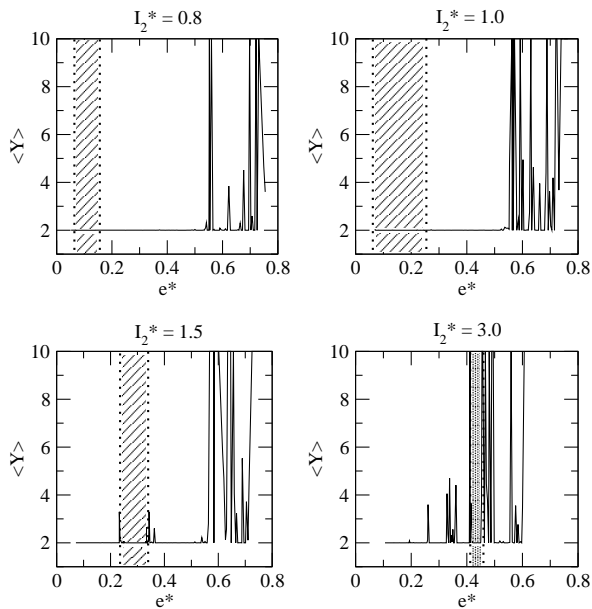}
\caption{Averaged MEGNO indicator $\langle Y \rangle$ for $200$ initial conditions with $\theta^*=0$ in the four dynamical maps shown previously. Regular orbits are characterized by $\langle Y \rangle \le 2$ while larger values are indicative of chaotic motion. Unstable orbits were given a value of $\langle Y \rangle=10$. In each plot the libration region is shown in shade.}
\label{fig5}
\end{center}
\end{figure}

Although most orbits in Figure \ref{fig3} appear regular, the map for $I_2^*=3.0$ shows some irregular trajectories which appear to indicate chaotic motion. To test this, we integrated 200 initial conditions for every value of $I_2^*$. Each had a different initial value of $I_1^*$ and the initial value of the resonant angle $\theta^*$ was taken equal to zero. Total integration time was $10^6$ years and, together with the equations of motion, we also calculated the MEGNO chaoticity indicator (Cincotta \& Simo 2000). Figure \ref{fig5} shows the values of averaged MEGNO number $\langle Y \rangle$ as a function of the mean-mean eccentricity for each map. Regular orbits yield $\langle Y \rangle \le 2$ while larger values are indicative of chaotic motion (see Cincotta and Simo 2000 for more details).
Unstable orbits, leading to an ejection of the particle in the integrated timespan, was assigned a value of $\langle Y \rangle=10$. The extent of the libration island is shown as a shaded rectangle. 

For low values of $I_2^*$ the chaotic region is restricted to high eccentricities and far from the secular resonance region. Thus, all commensurate trajectories are indeed regular, at least in the time interval analyzed. For $I_2^*=1.5$ the stochasticity of the resonance separatrix is clearly visible. However, for $I_2^*=3.0$ the chaos is much more extensive and covers the Jovian $\nu_\odot$ resonance zone. This helps explain why the dynamical map constructed for this value of $I_2^*$ showed irregular trajectories in the $(e^*,\theta^*)$ plane. A simple analysis of the output in osculating elements shows that the stochasticity is caused by the $6/1$ mean-motion resonance with the planetocentric orbit of the Sun. As shown recently by Hinse et al. (2009), the retrograde satellite region in the Jovian system is affected by several mean-motion commensurabilities. Even though they appear to be of high order, the interaction with the secular resonance gives origin to a noticeable chaotic behavior.

\section{Resonance Maps in the Saturn Satellite System}

Since the Jovian satellite system is tainted with chaos stemming from mean-motion resonances, we switched planets are analyzed the structure of the $\nu_\odot$ commensurability in Saturn. Beaug\'e \& Nesvorn\'y (2007) found that although the irregular satellite Narvi is not currently in a resonant configuration, the effects of the other outer planets will cause temporary trapping in the $\nu_\odot$ resonance in the future. Contrary to both Sinope and Pasiphae in the Jupiter system, the resonant behavior of Narvi appears to correspond to an asymmetric libration of $\theta^*$. 

\begin{figure}
\begin{center}
\epsfig{figure=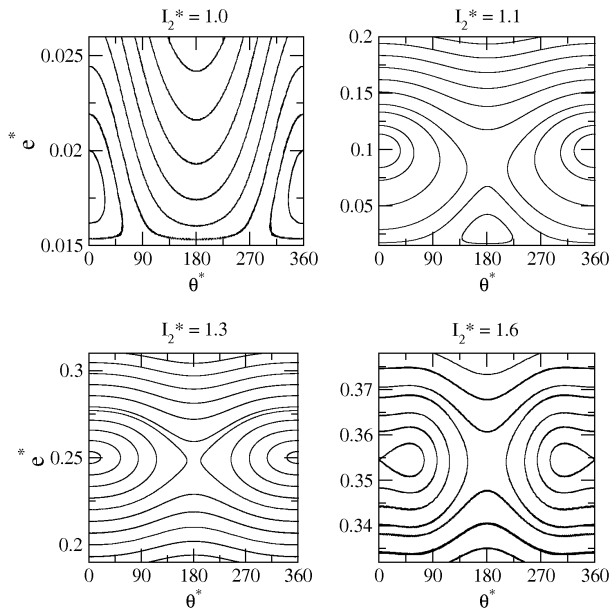,height=8cm,angle=0,clip=}
\caption{Structure of the $\nu_\odot$ resonance for initial conditions around the Saturn irregular satellite Narvi. Each plot shows filtered output, in the $(e^*,\theta^*)$ plane, of a series of numerical integrations with initial conditions leading to $a^* = 0.13$ AU and four different values of $I_2^*$ (in units of $10^{-5}$).}
\label{fig6}
\end{center}
\end{figure}

\begin{figure}
\begin{center}
\epsfig{figure=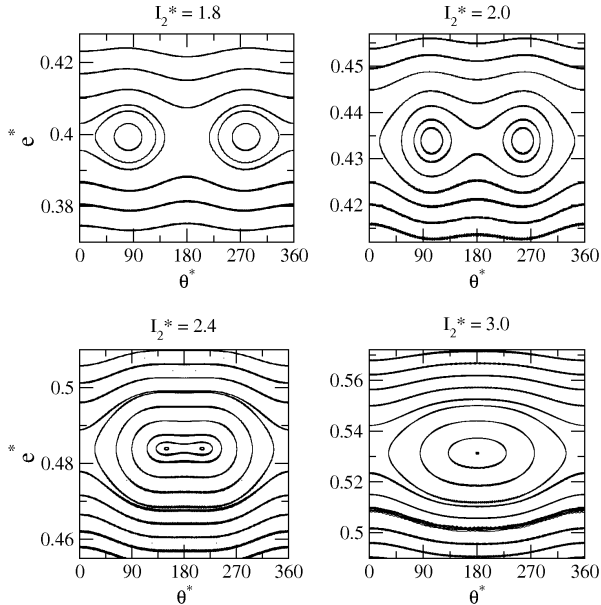,height=8cm,angle=0,clip=}
\caption{Same as previous plot, for larger values of $I_2^*$.}
\label{fig7}
\end{center}
\end{figure}

Taking the orbital elements of Narvi as a staring point, we constructed resonance maps for Saturn in an analogous manner as those constructed for Jupiter. In this case the proper semimajor axis was $a^* = 0.13$ AU (corresponding to $L^* = 1.04 \times 10^{-4}$), and we varied $I_2^*$ from $1.0$ to $3.0$ (in units of $10^{-5}$). For each value of $I_2^*$ we searched for initial conditions with different values of $I_1^*$. The results were filtered to eliminate the short-period terms associated to the mean anomalies and the variations due to the precession of the longitude of the node. Results are shown in Figures \ref{fig6} and \ref{fig7}. The first figure corresponds to lower values of $I_2^*$ and show significant similarities with the maps constructed for Jupiter (Figure \ref{fig3}). Libration at low eccentricities is associated to a libration around $\theta^*=0$, although for $I_2^*=1.6$ we note that the symmetric solution turns unstable and bifurcates into two islands of asymmetric libration. 

\begin{figure}
\begin{center}
\epsfig{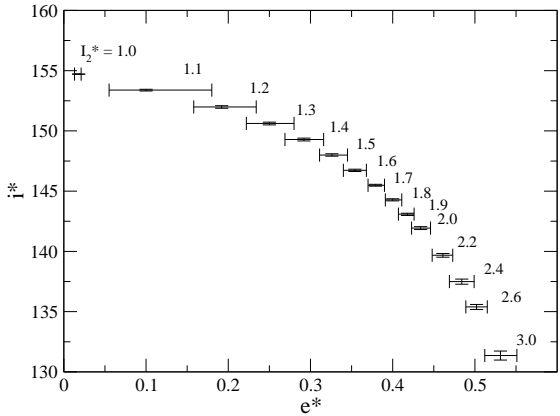}
\caption{For $L^*=1.04\times 10^{-4}$ (i.e. $a^* = 0.13$ AU) in the Saturn system, plot shows values of $i^*$ and $e^*$ corresponding to the center of the main libration island for different values of $I_2^*$. Error bars indicate the libration width in each orbital element.}
\label{fig8}
\end{center}
\end{figure}

The asymmetric structure appears more evident in the top-left hand plot of Figure \ref{fig7} ($I_2^*=1.8$). There are now two separatrix, an exterior homoclinic curve which stems from $\theta^*=180^\circ$ and surrounds the whole resonant domain, plus an interior separatrix centered around $\theta^*=0$ that encompasses the asymmetric libration centers. The order is inverted in the next plot, constructed for $I_2^*=2.0$, and now the critical curve associated to $\theta^*=0$ appears exterior in the resonance domain. As $I_2^*$ increases the equilibrium values of $\theta^*$ move towards $\theta^*=180^\circ$ until a new bifurcation is noted to $I_2^* \sim 2.8$. From this point onwards the libration returns to a symmetric solution, this time around $\theta^*=180^\circ$. 

It is interesting to note the change in the topology of the resonance as a function
of $I_2^*$ and, consequently, as function of the value of $e^*$ associated to the center of the libration domain. Previous studies of the $\nu_\odot$ commensurability only showed a libration around $\theta^*=0$ and, in the case of Narvi, a possible asymmetric libration. The complete structure is far more complex. Figure \ref{fig8} shows the values of $i^*$ and $e^*$ corresponding to the center of the main libration island for different values of $I_2^*$. The same general trend is observed as in Figure \ref{fig4}, where once again the resonance region appears restricted to mean-mean eccentricities below $e^* \sim 0.55$ for high values of the orbital inclinations. 

Finally, no significant chaos was found for these initial conditions, indicating that 
the $\nu_\odot$ resonance appears more stable for Saturn than for Jupiter. At least in part, this is due to the larger value of $a_\odot$ which pushes the dynamically significant mean-motion resonances much further from the planet.

\section{Satellite Migration}

So far we have only analyzed the resonant structure in Saturn for a given value of $L^*$, corresponding to $a^*=0.13$ AU. Since the construction of the dynamical maps is very time consuming, instead of repeating the calculations for other semimajor axes we have chosen to study the effects of a slow (adiabatic) migration on the satellite orbit. The planets were not affected by this migration and remained in fixed heliocentric orbits.

To simulate a smooth orbital decay we used the same procedure as described in Beaug\'e et al. (2006), which consists in including a Stokes drag-like exterior force in the equations of motion. This additional non-conservative perturbation can be modified to introduce any desired change in both semimajor axis and eccentricity. For all the integrations we adopted values such that the migration introduced no change in either the eccentricity or inclination, but only a slow change in $a^*$ with a characteristic time scale equal to $\tau_a=10^7$ years. This value is much larger than the librational period of $\theta^*$, thus guaranteeing the adiabatic approximation. 

\begin{figure}
\begin{center}
\epsfig{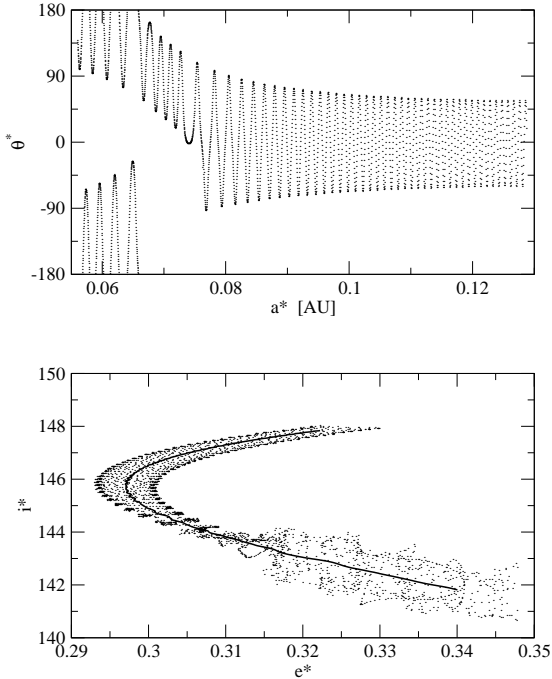}
\caption{Simulation of satellite orbital decay. Initial condition corresponds to a moderate amplitude $\theta^*=0$ libration with $I_2^*=1.5$ at $a^* = 0.13$ AU. Migration only affects the satellite semimajor axis with $\tau_a=10^7$ years. In the lower plot the thick black curve is the filtered output over the resonant angle.}
\label{fig9}
\end{center}
\end{figure}

Figure \ref{fig9} shows a typical example of an inward migration, where the initial condition corresponds to $a^* = 0.13$ AU and $I_2^*=1.5$ and a moderate amplitude of libration around $\theta^*=0$. The integration was continued until the final semimajor axis equaled $a^* \sim 0.06$ AU (closer to the planet than the real irregular moons), and the output was filtered in the same manner as before. The top graph shows the variation of the resonant angle as function of the semimajor axis, and the orbital evolution caused by the migration occurs from right to left. Although the particle remains trapped in the secular resonance throughout the integration, for smaller $a^*$ it switches from an oscillation around $\theta^*=0$ to an asymmetric libration, and finally, to a motion around $\theta^*=180$. The change in amplitude after each bifurcation is due to the discontinuity in the action at these points. The bottom plot shows the change in $i^*$ and $e^*$. during the migration. The thick black curve was obtained applying an additional filter in the numerical data to eliminated the variations with period similar to that of the resonant angle.

\begin{figure}
\begin{center}
\epsfig{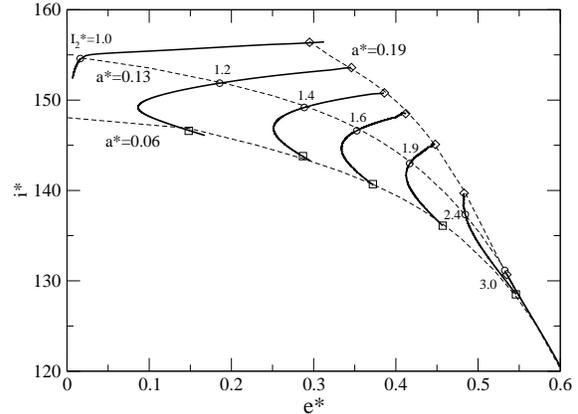}
\caption{Thick continuous curves show the families of center of libration as obtained from inward and outward migration starting from $a^*=0.13$ AU. The initial value of $I_2^*$ of each family is marked close to the initial position (empty circle). The inward migration was stopped at $a^* \simeq 0.06$ AU and the corresponding solution marked with an empty square. The outward migration was continued until $a^* \simeq 0.19$ AU. Here the final solution is marked with a diamond. Dashed curves show the interpolated families for the three values of $a^*$ discussed above.}
\label{fig10}
\end{center}
\end{figure}

The same procedure was repeated for initial conditions for other values of $I_2^*$ and extended to an outward migration of the satellite orbits. The resulting families in the $(i^*,e^*)$ plane are plotted Figure \ref{fig10}. Each thick continuous curve shows the filtered evolution of the resonant orbit during the orbital migration, and can be considered as the approximate location of the families of zero-amplitude solutions for different values of $a^*$. Empty rectangles show the location of the solutions for $a^*=0.06$ AU while diamonds correspond to $a^*=0.19$ AU. For larger semimajor axes the orbital behavior showed strong chaotic motion leading to an escape of the particles in short timescales. The initial conditions (i.e. center of libration for $a^*=0.13$ AU) are shown in empty circles. Dashed curves correspond to interpolated values calculated from the numerical solutions for all three semimajor axis.

The area inside the ``wedge''-shaped region gives a fair idea of the region covered by the $\nu_\odot$ resonance for retrograde orbits in the Saturn system, at least for the interval of $a^*$ analyzed. For high inclinations the libration domain occurs for low to moderate eccentricities, and the exact location depends strongly on $a^*$. However, as $i^* \rightarrow 90^\circ$, the resonance region shrinks until for $i^* < 130^\circ$ it is practically restricted to a single curve for all values of the semimajor axis.

\section{Scaling Law for Other Satellite Systems}

The structure of the $\nu_\odot$ resonance shows the same dynamical structure when computed for fictitious retrograde satellites in the Jovian and the Saturn system. Although constructed for different values planets and values of $L^*$, Figures \ref{fig3} (Jupiter) and \ref{fig6} (Saturn) appear basically the same, even if the values of $I_2^*$ and the $e^*$ of the librational centers are not equal. As a consequence, we wish to analyze whether there exists a simple transformation (or scaling) that allows us to deduce the main characteristics of the secular resonance around any planet, starting from the information obtained for Saturn. 

Switching planets introduces two main effects: (i) a change in the semimajor axis $a_\odot$ of the planetocentric orbits of the Sun (to a new value $a'_\odot$), and (ii) a change in the mass $m_p$ of the planet (to a new value $m'_p$). The outer planets also differ in eccentricity and inclination, although these parameters can temporarily be considered fixed. 

\begin{figure}
\begin{center}
\epsfig{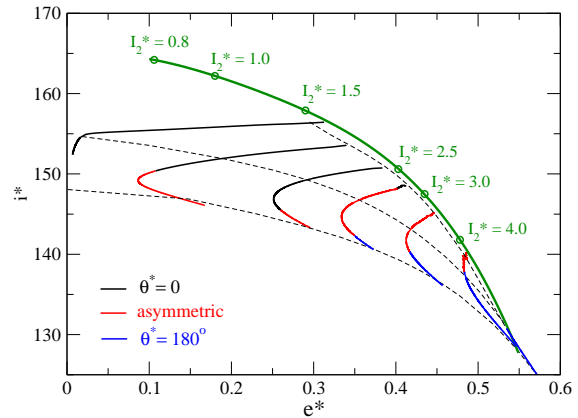}
\caption{Families of the zero-amplitude resonant solutions, as function of $a^*$, for retrograde orbits in the Saturn satellite system and six values of $I_2^*$. Black curves correspond to libration around $\theta^*=0$, red to oscillations around asymmetric libration centers and blue to libration around $\theta^*=180^\circ$. Dashed lines show interpolated curves for $a^*=0.06$, $a^*=0.13$ and $a^*=0.19$ AU. Thick green curve shows the interpolated family of solutions obtained for fictitious Jupiter satellites with $a^*=0.16$ AU (see Figures \ref{fig3} and \ref{fig4}).}
\label{fig11}
\end{center}
\end{figure}

Allowing for a change in $a_\odot$ is fairly straightforward. Since the secular Hamiltonian (averaged over short-period terms) only depends on the ratio $a/a_\odot$
(e.g. Murray \& Dermott, 1999), the semimajor axis of the planet only appears as a normalizing factor in the dynamical system. Thus, it should be expected that the dynamical structure calculated at $a^*$ in the first planet should correspond to the one located at $a'^* = \Gamma_a a^*$ in the second primary, where 
\begin{equation}
\label{eq6}
\Gamma_a = \frac{a'_\odot}{a_\odot} .
\end{equation}

A more complex task it trying to model the changes caused by differences in the planetary mass $m_p$. However, we can use an approximation of the restricted circular three body problem (RC3BP) known as Hill's Problem (Hill, 1878, 1886, see also Szebehely 1967, section 10.4). Hill's equations, originally constructed as the basis of a Lunar Theory, are a simplified version of the RC3BP in which the more massive primary (i.e. Sun) is considered to be infinitely far away but its gravitational effects are still felt in the system. In a rotating coordinate system with position vectors ${\bf r}$ centered in the smaller primary (i.e. planet), it is possible to introduce a scaling ${\bf r} \rightarrow {\bar \mu}^{1/3} {\bf r}$ which yields equations of motion that are independent of the masses (at least to smallest order in ${\bar \mu}$). We denote ${\bar \mu}$ as the mass ratio between primaries. Consequently, the complete dynamical behavior of the system can be deduced for any given value of ${\bar \mu}$ and extended to other masses just modifying the scaling parameter. 

Hill's approximation assumes very small mass ratios, motion close to the smaller primary and circular orbits between both massive bodies. Although these conditions are not rigorously satisfied in our case, its results may still be used as a first approximation. We may then introduce a second scaling factor $\Gamma_m$ defined as
\begin{equation}
\label{eq7}
\Gamma_m = \biggl( \frac{m'_p}{m_p} \biggr)^{1/3}
\end{equation}
which constitutes a new transformation in the coordinates due to changes in the planetary mass. Thus, given the complete change from one primary $m_p$ to another of mass $m'_p$ with different heliocentric semimajor axis, the original dynamical structure obtained at $a^*$ should be reproduced at a new value given by
\begin{equation}
\label{eq8}
a'^* = (\Gamma_a \Gamma_m) \, a^* .
\end{equation}

As a first test, Figure \ref{fig11} once again presents the families of zero-amplitude resonant solutions in the Saturn system (continuous black, red and blue curves) for smooth variations of $L^*$. However, now colors indicate the type of libration: black is used for motion around $\theta^*=0$, red for asymmetric librations and blue for oscillations around $\theta^*=180^\circ$. As already shown in Figure \ref{fig9}, migration can cause changes in the mode of libration. The dashed curves show the interpolated location of the centers for $a^*=0.06$, $a^*=0.13$ and $a^*=0.19$ AU. Thus, for any other value of the proper semimajor axis it is possible, at least qualitatively, to estimate the locus of zero-amplitude solutions in the $(e^*,i^*)$ plane and the corresponding type of libration. 

As an example, let us consider the resonant maps constructed in the Jovian system for $a^*=0.16$ (Figure \ref{fig3}). The locus of resonant centers shown in Figure \ref{fig4} can be interpolated and is shown in Figure \ref{fig11} as a thick continuous green curve. The individual values are marked by open circles with the corresponding values of $I_2^*$ placed adjacently. Applying the scaling law (\ref{eq8}) we find that in the Saturn system this curve should correspond to $a'^*=0.197$ AU, placing it barely above the upper dashed line. Not only does the location of the green curve shows a very good correlation with the prediction of this simple scaling law, but there is also an agreement between the predicted type of resonant mode and the dynamical maps shown in Figure \ref{fig3}. It is important to mention that the eccentricity adopted for each planetary body were equal to their individual present values. 

\begin{figure}
\begin{center}
\epsfig{figure=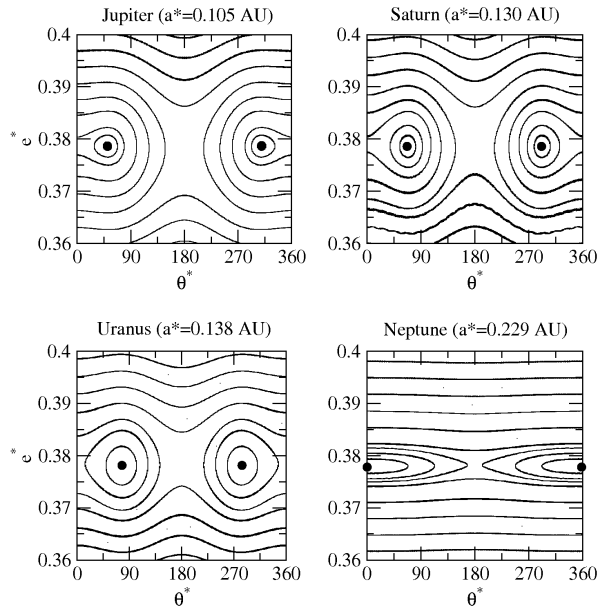,height=8cm,angle=0,clip=}
\caption{Resonant maps, around all giant planets, for initial conditions leading to same location of the libration center (black circle) in the $(e^*,i^*)$ plane. Values of the proper semimajor axes are given on the top of each frame. While the first three maps are very similar, the low eccentricity of Neptune prevents the appearance of the asymmetric domain.}
\label{fig12}
\end{center}
\end{figure}

In conclusion, the libration centers (both location and type) of fictitious satellites around Jupiter can be deduced from their counterparts in the Saturn system, simply applying the scaling law (\ref{eq8}) to the original proper semimajor axis $a^*$ and analyzing the curve corresponding to the new $a'^*$ in Figure \ref{fig11}. However, we still need to analyze the structure of the resonance lobes in more detail and extend the study to other planetary masses. Results are shown in Figure \ref{fig12}, where we present four dynamical maps, each for fictitious satellites around a different giant planet. Starting from Saturn, the values of $L^*$ and $I_2^*$ adopted for the other frames were deduced from the scaling law. The value of $I_2^*$ was chosen close to the onset of the asymmetric librations, and thus constitutes a region particularly sensitive to changes in the dynamical system. 

If the planetary eccentricities  were taken the same, the resonant structure around all four outer planets would be identical. However, since the value of $e_\odot$ is different for each primary, the resonant structure must reflect this change. The location of the libration center in $e^*$ and $i^*$ appears practically invariant, with $e^* \simeq 0.379$ in all four plots. Moreover,  Jupiter, Saturn and Uranus have very similar orbital eccentricities, and therefore the type of libration (i.e. asymmetric) is maintained and the resonant domain has similar sizes. However, since $e_{\rm Sat} > e_{\rm Ura} > e_{\rm Jup}$, the libration lobe surrounding the asymmetric solution is larger in Saturn than in Uranus, while Jupiter presents the smallest region of the three.

Although the differences in topology of the resonant domain around the first three planets are not very significant, the map constructed for Neptune shows no asymmetric lobe and the libration occurs around $\theta^*=0$. This, however, may be understood in terms of the planetary eccentricity $e_{\rm Nep} = 0.008$, much smaller than the other giant planets. Let us recall the first-order analytical model $F_1^*$ presented by Yokoyama et al. (2003) and summarized in equation (\ref{eq2}). Writting the coefficients $A_i$ explicitly in terms of the perturber's eccentricity $e_{\odot}$, and retaining only smallest orders, we can express this function as:
\begin{equation}
\label{eq10}
F_1^* = C_0 + C_1 e_\odot \cos{(\theta^*)} + C_2 e_\odot^2 \cos{(2\theta^*)},
\end{equation}
where the new coefficients $C_i=C_i(I_1^*;L^*,I_2^*,m_p)$ are now independent of $e_\odot$. Although we have mentioned that this model is not a reliable representation of the dynamics of the $\nu_\odot$ resonance, the imprecision is only due to the values of the coefficients $C_i$. The functional form of the resonant Hamiltonian stems from the D'Alembert rules of any expansion of the perturbing potential, which specifies that, for any integer $k$, the terms associated to a $k$-order harmonic of $\theta^*$ must be proportional to $e_\odot^k$. Thus, any other Hamiltonian function, whatever its origin, will have the same form as (\ref{eq10}). 

Libration centers are specified by the conditions $\partial F^*_1/\partial I_1^*=\partial F^*_1/\partial \theta^*=0$. In particular, the second condition explicitly reads
\begin{equation}
\label{eq11}
\sin{(\theta^*)} \biggl( C_1 + 4 C_2 e_\odot \cos{(\theta^*)} \biggr) = 0.
\end{equation}
This equation has two trivial solutions ($\theta^*=0$ and $\theta^*=180^\circ$) which correspond to symmetric librations. However, if 
\begin{equation}
\label{eq12}
4 \, |C_2| \, e_\odot > |C_1|,
\end{equation}
the contribution of the second harmonic in the resonant angle is larger than the first, and asymmetric librations are possible (see Beaug\'e 1994 for a similar analysis for exterior mean-motion resonances in the asteroidal problem). Since $C_1$ and $C_2$ vary differently as function of $L^*$ and $I_2^*$, we can explain why asymmetric librations are found for some values of the inclination and not for others. However, condition (\ref{eq12}) depends linearly on the planetary eccentricity. Thus, if $e_\odot$ is sufficiently small, asymmetric solutions are not possible and all librations are of the symmetric type. This appears to be the case of the Neptune satellite system, at least for the values of $L^*,I_2^*$ analyzed. 

\begin{figure}
\begin{center}
\epsfig{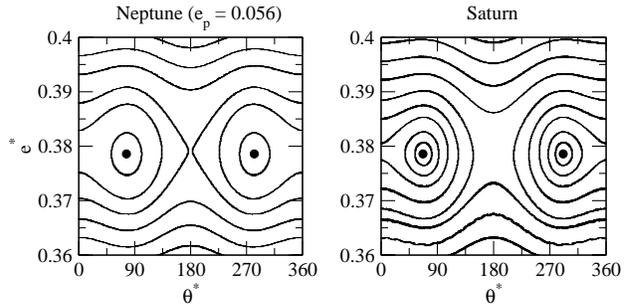}
\caption{Same as Figure \ref{fig12}, comparing the resonant maps for Neptune (left) and Saturn (right). However, now the eccentricity of both planets were considered the same, and equal to the osculating value of Saturn. The resonant structure in both frames is identical.}
\label{fig13}
\end{center}
\end{figure}

As a final test, Figure \ref{fig13} shows a comparison of the resonant structure of two systems: the plot on the right reproduces the dynamics around the asymmetric libration lobe in Saturn (top-right frame of Figure \ref{fig12}), while the left-hand graph was obtained for a fictitious Neptune in which the planet's eccentricity was elevated to the same value as Saturn. As expected, the resonant structure is now equivalent, confirming that the absence of asymmetric librations in Figure \ref{fig12} is in fact due to the planet's quasi-circular orbit.

\section{Conclusions}

We have presented a numerical study of the structure of the $\nu_\odot$ secular resonance for retrograde satellites around the giant planets. This commensurability is associated to a libration of the secular angle $\theta=\varpi-\varpi_\odot$, where $\varpi = \Omega-\omega$ is the longitude of pericenter of the irregular moon and $\varpi_\odot$ the corresponding angle for the planetocentric orbit of the Sun. We have shown that the resonant domain presents a diversity of libration modes, including both symmetric and asymmetric librations. Each mode appears restricted to certain intervals of the mean-mean inclination $i^*$ of the satellite. Librations around $\theta=0$ occur for large values of $i^*$ (typically $i^* \gtrsim 150^\circ$), while oscillations around $\theta=180^\circ$ seem to dominate for $i^* \lesssim 140^\circ$. Asymmetric solutions appear at intermediate inclinations, although planets with very small eccentricities (such as Neptune) may not contain asymmetric solutions at all. Calculating the mean-mean inclination for real irregular satellites shows a good agreement with this predictions. Pasiphae ($i^*=148^\circ$), Sinope ($i^*=157^\circ$) both display librations around $\theta=0$, while the resonant angle of Narvi ($i^*=141^\circ$) is in an asymmetric mode. 

Simulating a smooth orbital migration of satellite orbits have shown that the resonant lock is maintained as long as the adiabatic condition is met; in other words, the characteristic timescale of the orbital variation is much longer than the libration period. We have used these results to map the location and type of libration centers as a function of the proper semimajor axis, thus giving a more global view of the region of the phase space dominated by the $\nu_\odot$ commensurability. 

Finally, we presented a simple scaling law that can be used to relate the resonant structure in any the satellite system of planetary mass. Since Jupiter, Saturn and Uranus have similar orbital eccentricities, it is expected that the structure of the secular resonance should also be very similar, except for a scaling in the semimajor axis of the irregular moons. Neptune, however, has an almost circular orbit, which inhibits the formation of asymmetric libration points. For satellites around this planet, it is expected that symmetric solutions should dominate. 

From the scaling law, it can be shown that a change in the semimajor axis of the satellite is equivalent to a similar one on the planet. In other words, the same results would obtained if the heliocentric semimajor axis of the planet suffered a slow migration while the planetocentric orbit of the moon was kept fixed. Although we have only done a few simulations with a fast (non-adiabatic) migration, it appears that the resonant configuration is not so easily preserved, and any originally librating body was usually expelled from the resonance. However, additional simulations are required to confirm is this behavior is general.

In this paper we have considered only the restricted three-body problem and, consequently, neglected the gravitational effects of the other giant planets. This is indeed an approximation, since these additional perturbations may cause significant orbital variations (e.g. Carruba et al. 2004, \'Cuk \& Burns 2004). In particular, the temporary nature of the $\nu_\odot$ resonance lock of real satellites is due to the gravitational effects of the other planetary bodies. In consequence, the results presented here must be viewed only as a starting ground for the complete secular behavior for retrograde irregular moons. Surely the full picture will be far more complex.

\section*{Acknowledgments} 
This work has been supported by the Cordoba National University (Secyt-UNC) and the Argentinian Research Council (CONICET).

\end{document}